Research Article                                                                                                **PHYSICS**

# Degradation of Cyanobacterium *Nostoc muscorum* via Air, Oxygen, and Nitrogen low temperature plasmas

**Samaa Salah[1], Mohamed Mokhtar Hefny[2], Nabil M. El-Siragy[1], Abdelhamid Elshaer[3], Eithar El-Mohsnawy[4] & Mohammed Shihab[*1]**

[1]Department of Physics, Faculty of Science, Tanta University, 31527 Tanta, Egypt
[2]Engineering Mathematics and Physics Department, Faculty of Engineering and Technology, Future University in Egypt, Cairo 11835, Egypt
[3]Department of Physics, Faculty of Science, Kafrelsheik University, Kafr-Elsheik, Egypt
[4]Department of Botany and Microbiology, Kafrelsheik University, Kafr-Elsheik, Egypt

**Introduction**

Cyanobacteria are prokaryotic organisms capable of converting light energy into chemical energy through photosynthesis (**Sinha and Häder, 2008**). These organisms have numerous potential applications, including bioethanol production, food colorings, dietary supplements, and raw materials (**Pathak el al., 2018**). Cyanobacteria can be found in a variety of environments, including freshwater and terrestrial habitats, as well as in extreme environments such as hot springs, hypersaline water, polar regions, and low-oxygen environments (**Stal and Tamulonis, 2011**). During the light reaction, chlorophyll, carotenoids, and phycobilins absorb light energy and convert it into high-energy compounds like ATP and NADPH, while also producing oxygen. Cyanobacteria have played a vital role in enriching the earth's atmosphere with oxygen over many centuries (**Whitton, 2012**). In anaerobic conditions, some cyanobacteria can fix atmospheric nitrogen through heterocysts, producing ammonia, nitrites, or nitrates, which are essential for the production of proteins and nucleic acids. The main pigments in cyanobacteria include peridinin (**Fuciman et al., 2011; Zigmantas et al., 2004; Polivka et al., 2007**), $\beta$-carotene, nostoxanthin, xanthophyll, aloxanthin, echinenone, myxoxanthophyll, canthaxanthin, oscillaxanthin, zeaxanthin, and scytonemin (**Proteau et al., 1993; Garcia-Pichel and Castenholz, 1991; Colyer et al., 2005; Fresnedo et al., 1991; Bergamonti et al., 2011; Balskus et al., 2011**). Carotenoids play a crucial role in energy dissipation and energy transfer to the reaction center. They are organic soluble pigments found in lipophilic environments within the thylakoid membranes (**Melendez-Martinez et al., 2007**). Cyanobacteria exist in various forms, including unicellular and filamentous, and *Nostoc muscorum* colonies can consist of filaments, sheets, or even hollow spheres (**Dodds et al., 1995**). Unfortunately, cyanobacteria can also form harmful algal blooms, which can disrupt aquatic ecosystems and cause intoxication in wildlife and humans due to the production of toxins such as microcystins, saxitoxin, and cylindrospermopsin. These blooms are increasing in frequency and magnitude globally, posing a serious threat to both aquatic environments and public health (**Paerl and Otten, 2013**). In addition to producing oxygen, cyanobacteria also produce toxins that can have harmful effects on both humans and animals.

Traditional methods of water treatment based on filtration techniques are insufficient in eliminating bacteria, viruses, and harmful organic substances from water. Furthermore, the use of chemicals such as chlorine can result in the formation of harmful chemical compounds and pose health risks to consumers. Plasma water treatment is a promising and growing field for the removal of organic and toxic contaminants from water **(Zeghioud et al., 2020; Magureanu et al., 2018; Ukhtiyah et al., 2023; Takeuchi and Yasuoka, 2020)**. By applying high voltage to the water, plasmas are created, resulting in highly reactive free radicals that can attack chemical and biological pollutants in the water. Plasma treatment is a cost-effective, energy-efficient, and low-maintenance method for disinfecting water and can be used in conjunction with conventional filtration processes to remove toxic and organic compounds from industrial wastewater and water sanitation. This type of plasma is referred to as non-thermal plasma, cold plasma, or non-equilibrium plasma, as it is not in thermodynamic equilibrium. The electron temperature is much hotter than the temperature of ions and neutrals, resulting in a significantly different velocity distribution for electrons compared to ions. The degree of ionization is also very small, with only one electron and one ion generated for every $10^6$ neutral particles **(Takeuchi and Yasuoka, 2020; Lieberman and Lichtenberg, 1994; Chabert and Braithwaite, 2011; Makabe and Petrovic, 2006)**. Although the electrons are heated up to a few electronvolts (1 electronvolt = $11600°K$), the rest of the gas remains neutral at room temperature. A type of common non-thermal plasma is found in fluorescent lamps, where electrons reach a temperature of 20,000 Kelvin, while the rest of the gas, ions, and neutral atoms remain barely above room temperature. As a result, the bulb can be touched with the hands while it is operating. This provides the opportunity to utilize low-temperature plasma in industry without causing thermal damage to biological samples and substrates **(Laroussi, 2018)**. There are several plasma devices available, such as capacitively coupled plasmas **(Shihab et al., 2022; Shihab and Mussenbrock, 2017; Shihab, 2018; Shihab et al., 2021)**, inductively coupled plasmas **(Laroque et al., 2022)**, dielectric barrier discharges **(Abd El-Latif et al., 2024)**, and atmospheric plasma jets **(Bekeschus et al., 2017; Hefny et al., 2016)**. These devices aim to generate low-temperature plasmas with varying densities and average temperatures close to room temperature. When electrons with a few electronvolts collide with the background gas, they produce chemical species that cannot be generated using traditional chemical methods. In these discharges, tens of chemical reactions occur based on the gas composition and pressure, as well as plasma temperature and density (**Shihab et al., 2021; Hamza et al., 2023; Abdel-Wahed et al., 2022**). There are two scenarios for the interaction of plasma with cyanobacteria. In the first scenario, the accelerated ions and high-speed neutral particles within the plasma sheath collide with the bacteria, potentially causing them to crack. In the second scenario, reactive oxygen and nitrogen species (RONS) generated by highly energetic electrons break the chemical bonds, leading to the formation of new structures that damage the bacteria. Both scenarios result in the destruction of the bacteria within a short period of time. The objective of this manuscript is to investigate the efficiency of air, oxygen, and nitrogen plasma in degrading cyanobacterial cells for use in wastewater treatment.

**Materials and Methods**
**Cyanobacterial sample**
Type of the used bacteria is Phylum: Cyanobacteria, Class: Cyanophyceae, Subclass: Nostocophycidae, Order: Nostocales, Family: Nostocaceae, Genus: Nostoc, species: muscorum.

*Nostoc muscorum* was obtained from microbial biotechnology lab, Faculty of Science, Kafrelsheikh University. Cyanobacterium was cultivated on BG-11 medium, sterilized air bubbling and white light intensity of 80 µmol.m$^{-2}$.s$^{-1}$.

**Plasma generation**
The plasma is generated via a dc discharge. Two circular electrodes with a radius of 2.5 cm. The gap between the two electrodes is 2.5 cm. The applied potential is 600 V, and the gas pressure is 0.2 mbar. The low pressure is achieved using a rotary pump. In each treatment, 1g of Cyanobacteria is mounted on Fluorine Tin Oxide-coated glass substrate, which is a transparent conductive substrate. The chamber is made of stainless steel and the two electrodes are made of copper. The plasma spectrum is measured during the treatment through a Pyrex glass, which absorbs radiation in the ultraviolet range below 300 nm. Figure (1) shows a schematic of the experiment setup.

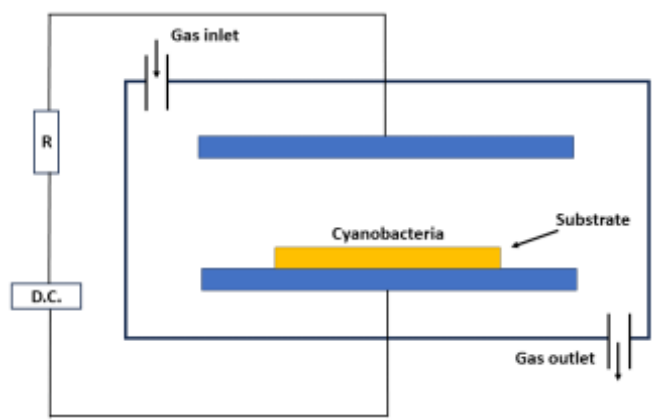

**Fig. (1):** Schematic of the experiment setup.

The plasma spectrum is measured during the plasma treatment of Cyanobacteria. Three gases are used: Air, Oxygen, and Nitrogen. The plasma species are displayed in Fig. (2). It is mandatory to emphasis that the Cyanobacteria constituents contribute to the measured spectrum.

**Results**

**Optical emission measurement**

The optical emission spectrum (OES) is measured during the plasma treatment of Cyanobacteria, where three gas conditions were used: air, oxygen, and nitrogen. The OES during the treatment can be seen in Fig. (2), where many spectral lines can be detected. These spectral lines are corresponding to RONS, which produce biological effects such as: OH- excited species at 314 nm, the second positive system of $N_2$ from 314 to 380 nm, the first negative system of $N_2$ from 390 to 470 nm, and the first positive system of $N_2$ around 545 nm. Moreover, some lines such as 357 and 391 nm can be attributed to both nitrogen and oxygen. Even though plasma is the main source of the measured OES, Cyanobacteria contribute to the measured spectrum since it is a source of oxygen and nitrogen as mentioned before **(Akter et al., 2020; NIST, 2024)**. The plasma generated high energetic electrons are responsible for generating high densities of metastables and radicals, which are the main reason of the plasma induced chemistry. Therefore, it is important to estimate their density and temperature to ensure that they are in the reported range of cold plasma. The electron density ($n_e$) was estimated by the Stark broadening method, see Fig. (3a). Using Stark broadening method, $N_2$ 391 nm was chosen, where we calculated its full width at half maximum (FWHM) of the Stark broadening ($\lambda_{Stark}$), and the electron density was estimated through the Lorentzian fitting of the line spectra with the help of Eq. (1). However, it should be mentioned that this method is valid only for estimation of the electron density, and for accurate measurements other broadenings of the spectral line such as instrumental broadening van der Waals broadening, and Doppler broadening should be taken into consideration **(Nikiforov et al., 2015)**.

$$\lambda_{Stark} = 2 \times 10^{-11} n_e^{2/3} \quad (1)$$

It was found that $\lambda_{Stark} \approx 1.7\ nm$, and consequently $n_e$ of approximately 2.48 x 10$^{16}$ cm$^{-3}$. Moreover, the electron excitation temperature was also estimated using Boltzmann plot method. In this method, several emission lines are used to estimate electron temperature according to the following equation:

$$log\left(\frac{(I_{nm}(2)/I_{nm}(1))/\lambda_{nm}}{g_m(2)A_{nm}(2)}\right) = \frac{-0.625}{T}E(2) + C \quad (2).$$

Where, $I_{nm}(1)$, $I_{nm}(2)$ are the emission intensities of the base line and the emission intensity of upper level, $g_m(2)$ is the statistical weight of the upper level of the transition, $A_{nm}(2)$ is the atomic transition probability, $\lambda_{nm}$ is the wavelength of the emitted upper level (nm), $E(2)$ is the excitation energy of the emitted upper level (cm$^{-1}$), T is the electron temperature (°K), and C is a constant. Figure (3b) represents the Boltzmann plot of Eq. (2), where we chose five spectral lines of N$_2$ (see Table 1) for this plot, E represents the horizontal axis, and $log\left(\frac{(I_1/I_0)/\lambda_{nm}}{A_1 g_1}\right)$ represents the vertical axis. The electron temperature can be calculated from the slope together with Eq. (2), and it was about 2.4 eV (27949 °K), which lies in the normal range of electron temperature for cold plasma **(Magureanu et al., 2018, Chen and Li, 2015)**. It is really important to emphasis that the temperature of all samples and the electrodes are at room temperature.

**Table (1):** Spectroscopic Data of N II

| $\lambda$ (nm) | Spectrum | E$_{exc}$ (cm$^{-1}$) | gm. Am (s$^{-1}$) |
|---|---|---|---|
| 337 ($\lambda_0$) | N II | 317395.2 | 7.62 |
| 391 | N II | 228731.81 | 1.21 |
| 399 | N II | 174212.03 | 6.1 |
| 405 | N II | 168892.21 | 0.0000432 |
| 545 | N II | 188909.17 | 0.295 |

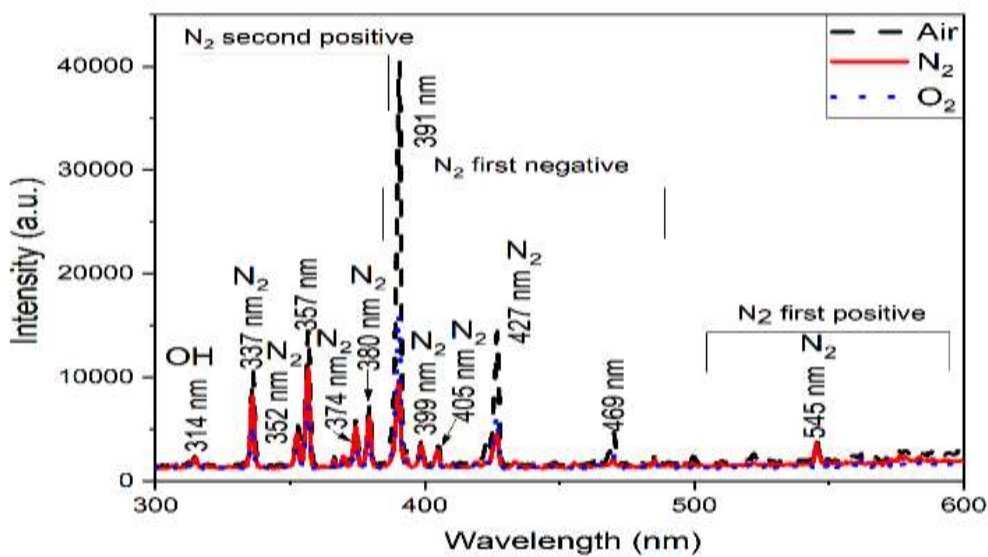

**Fig.(2):** The plasma OES during the treatment.

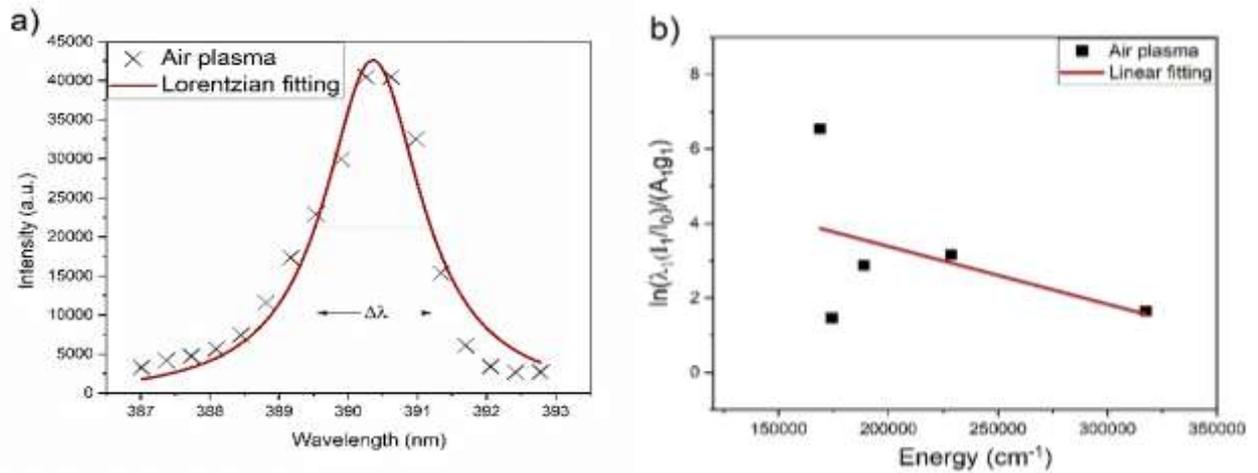

**Fig. (3):** (a) the Lorentzian fitting of N2 391 nm line profile for measuring electron density using Stark broadening method, and (b) Boltzmann plot for air plasma to estimate the electron temperature.

**Cyanobacteria growth before and after treatment**

**Experimental procedures**

*N. muscorum* is retrieved from the vibrating incubator where it is stored. Subsequently, the samples are positioned. *N. muscorum* was then collected from the medium using a centrifuge (5000g for 15 min), with *N. muscorum* was retained and the media discarded. *N. muscorum* were placed on substrates, and their weights are duly recorded. After this step, *N. muscorum* was introduced into the plasma chamber, positioned between two electrodes, and the device is sealed prior to plasma generation. *N. muscorum* was exposed for different durations for each gas. After exposure to the plasma, UV and Raman measurements were taken, and samples were cultivated in its dedicated medium to assess the impact of the plasma.

**Cyanobacteria growth behavior**

*N. muscorum* was exposed to plasma for different durations time (15, 30, 45 and 60 min), followed by monitoring the *N. muscorum* growth behavior. After the specified exposure period, the sample was transferred to BG-11 medium for cultivation, and the results were awaited. Cyanobacterium was observed for a week to note the effects of the plasma and any changes that occurred during this time. It was observed that the *N. muscorum* exposed to oxygen and nitrogen plasma for an hour showed a comlplete disappearance, where the medium appeared completely clear. This indicates that plasma has effectively eliminated the growth of *N. muscorum*. Meanwhile, *N. muscorum* exposed for 15 minutes exhibits less impact from the plasma. This indicates that the longer the sample is exposed to the plasma, the inhibiting of the growth of *N. muscorum*. Counting Cyanobacterial colonies using human eye is not an easy task. Unfortunately, colonies overlap and usually differ in color, shape, size, and contrast. However, the number of colonies was still a good parameter to determine the rate of growth of *N. muscorum*. The number of colonies was estimated employing NIST's Integrated Colony Enumerator (NICE**)** **(Clarke et al., 2010)**. The images of agar plates are imported by NICE, each image was divided into a large number of pixels. Pixels with the same color are counted giving a column in a histogram. This is one of the advantages of artificial intelligence, where images can be converted into quantitative numbers. The color of the pixel reflects a natural process or a parameter. If the histogram contains one column, then the image was homogeneous and contains one parameter. Different columns mean different parameters in the image. The height of a column in the histogram reflects the degree of dominant of its parameter. The idea is used widely in radiomics; where, the spatial distribution of colors and pixel interrelationships of tumor images produce useful quantitative analysis as the tumor size, homogeneity, and entropy **(Houseni et al 2021; Mahmoud et al., 2021a; Mahmoud et al., 2021b)**. Since colonies are usually dark, then dark pixels are counted, and this yields estimation to the number of bacteria colonies.

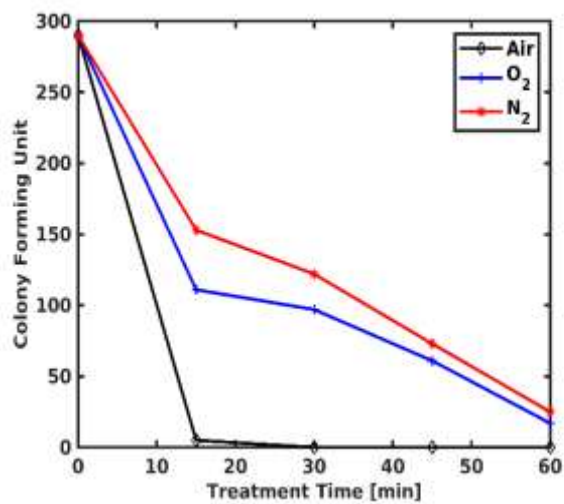

**Fig. (4):** The colony forming unit as a function of the plasma treatment time was repeated three times providing results with an error less than 2%.

Figure (4) displays the effect of plasma on cyanobacteria. It is clear from figure (4) that air plasma is very effective. It is able to kill all cyanobacteria after 15 min treatment. Oxygen or nitrogen plasma need more time a complete removal of cyanobacteria. The air plasma, which consists of oxygen and nitrogen, was more effective than oxygen plasma or nitrogen plasma. This can be attributed to high concentration of RONS in case of air plasma than oxygen or nitrogen plasma, which appears clearly from their OES (see Fig.2).

**Light absorption**

The absorption spectra (400 – 700 nm) of *N. muscorum* colonies were measured using V-630 UV-Vis Spectrophotometer from Jasco. It is a compact instrument with many features as it covers ranges from 190 to 1100 nm and high-speed scanning up to 8000 nm/min. In Figure (5) the absorption spectrum of control sample (without any plasma treatment), and samples treated with air plasma for 15 min, 30 min, 45 min, and 1 hr. The control sample shows absorption peaks 1-4 that refer to Chl a, Carotenoids, phycobilins and Chl a, respectively.

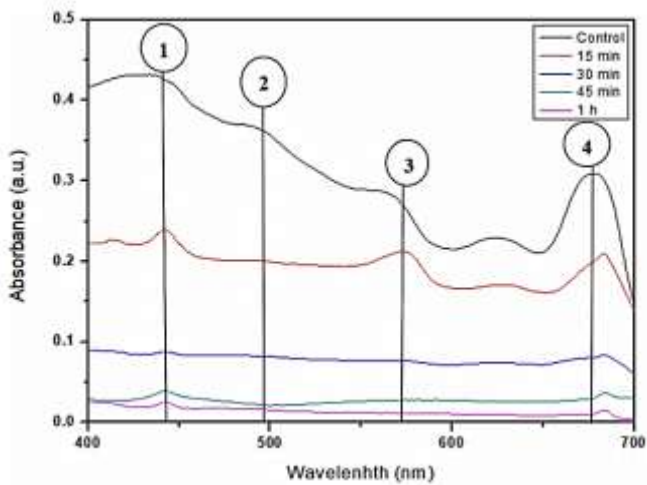

**Fig. (5):** The light absorption of Cyanobacteria before and after plasma treatment. The absorption of Chl *a* (1,4); Carotenoids (2); and phycobilins (3).

Due to the interaction of cyanobacteria with plasma causing a degradation of its cell structure, many chemical bonds are broken, and different chemical reactions take place. Therefore, the Cyanobacteria pigments disappear due to the interaction with plasma. The amplitude of light absorption peaks decreases by increasing the treatment time. The light Absorption illustrates the impact of plasma on *N. muscorum*. Each pigment in *N. muscorum* has a specific wavelength.

**Ramman spectrum of treated cyanobacteria**

Micro-Raman spectroscopy measurements were performed using an alpha 300 R confocal microscope (WITec GmbH). A 100X objective microscope was used to focus the laser beam on the sample to a size of ~300 nm (Diffraction Limit by 532 nm). A 532 nm green laser excitation was used. The spectrometer wavenumber scale was calibrated using the 521 rel.cm$^{-1}$ line of Si (this method is included but WITec uses the atomic lines of a halogen lamp for calibration, this method is more accurate). Raman spectroscopy via Witec alpha-300 was used to track the change of chemical composition of the Cyanobacterium before and after the plasma treatment. Raman modes for Cyanobacteria have been reported in the literature (**Pinzaru et al., 2016; de Olveira et al., 2015; Allakhverdiev et al., 2022; Němečková et al., 2023; Storme et al., 2015; Cui et al., 2023**). Characteristic bands around 1516, 1156 and 1006 cm$^{-1}$ have been measured due to beta-carotene in protein-carotenoid-complex in cyanobacteria strains. Due to the carotenoid, the cyanobacteria could be characterized instantly in vivo under the 532 nm and 1064 nm excitation. Raman Spectroscopy is not a destructive technique. Therefore, the growing of Cyanobacteria could be tracked based on their chemical composition and carotenoid accumulation. Here, our control sample provides the Raman peaks confirming the existence of Cyanobacteria (see Fig. 6). After the plasma treatment the Raman peaks disappeared and replaced with new peaks, indicating the degradation of the cyanobacteria and the formation of new compounds and molecules. The new peaks depend on the treatment time.

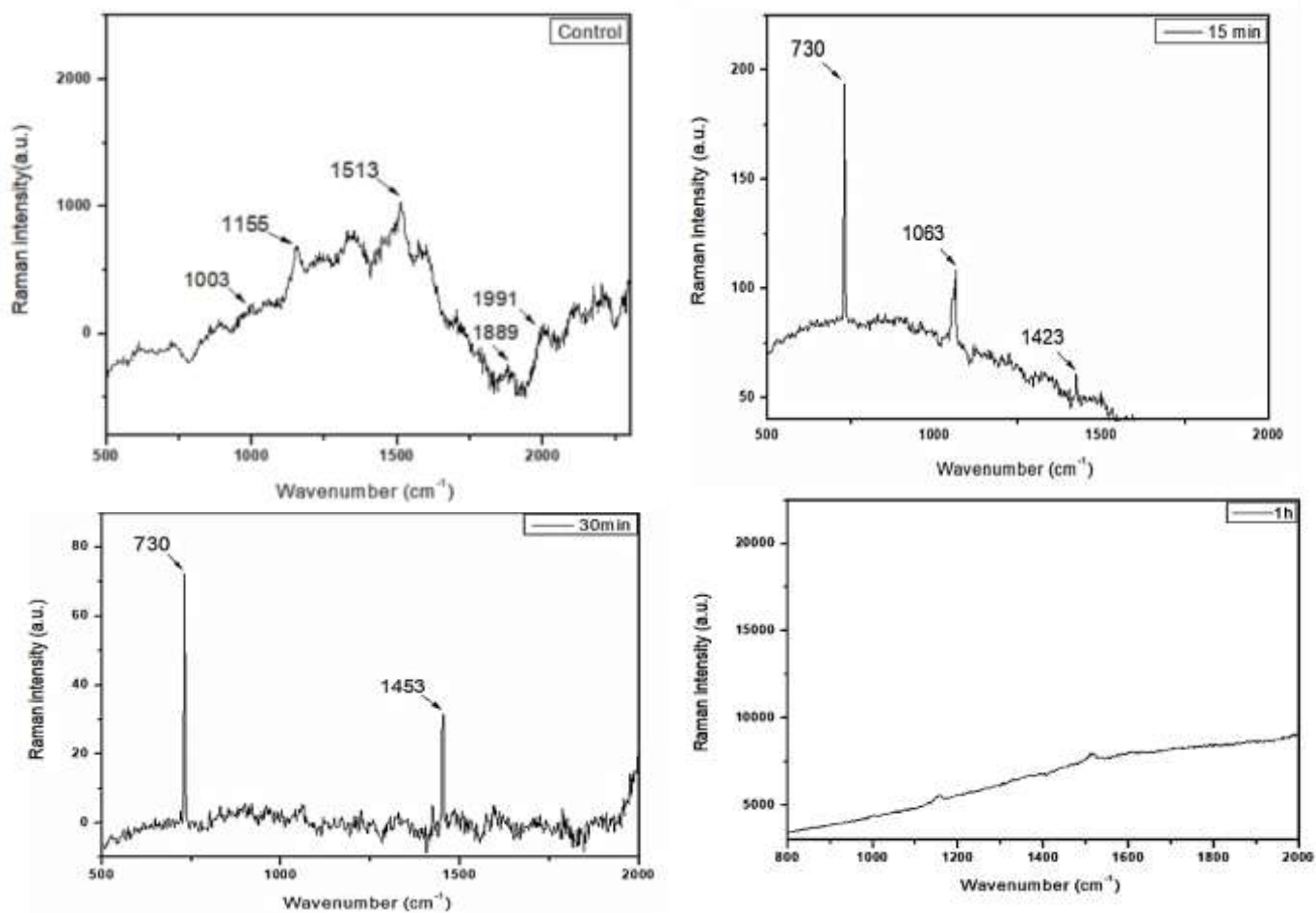

**Fig. (6):** Raman spectrum of Cyanobacteria before and after plasma treatment.

**Conclusion and future work**

Cyanobacterium *Nostoc muscorum* was exposed directly to the air, oxygen, and nitrogen plasma. Air plasma was the most powerful condition compared to oxygen and nitrogen plasma, where it achieved a complete degradation of *N. muscorum* within only 15 min, while oxygen and nitrogen plasma needed at least 1 hour for the complete degradation. The OES of air, oxygen and nitrogen plasma were measured and many plasmas generated species such as $N_2^+$, $N^+$, $OH$ were detected. In general, the highest concentrations of species were recorded for air plasma. The UV absorption and the characteristic Raman peaks of cyanobacteria pigments gradually disappeared with the plasma treatment time until a complete disappearance. Further work is needed to fully understand the effect of plasma treatment on Cyanobacterium *Nostoc muscorum*, especially in wastewater real samples.

## Author Contributions
All authors contributed to the study conception and design equally. All authors read and approved the final manuscript.

## Data availability
The datasets used and/or analysed during the current study are available from the corresponding author on reasonable request.

**تحلل السيانوبكتيريا *Nostoc muscorum* عبر البلازما منخفضة الحرارة باستخدام الهواء والأوكسجين والنيتروجين**

سما صلاح[1]، محمد مختار حفني[2]، نبيل محمد السراجي[1]، عبدالحميد الشاعر[3]، إيثار المحسناوي[4]، محمد شهاب[1]

[1] قسم الفيزياء، كلية العلوم، جامعة طنطا.
[2] الفيزياء والرياضيات الهندسية، كلية الهندسة والتكنولوجيا، جامعة المستقبل.
[3] قسم الفيزياء، كلية العلوم، جامعة كفرالشيخ.
[4] قسم النبات والميكروبيولوجي، كلية العلوم، جامعة كفرالشيخ.

السيانوبكتيريا هي كائنات ميكروبية بدائية النواة تمتلك القدرة على تحويل الطاقة الضوئية إلى طاقة كيميائية من خلال عملية التمثيل الضوئي. تعرض هذه الكائنات العديد من التطبيقات المحتملة، بما في ذلك إنتاج الإيثانول الحيوي، وتكوين الأصباغ الغذائية، وتطوير المكملات الغذائية، وتوفير المواد الخام. تُظهر السينوبكتيريا توزيعًا بيئيًا واسعًا، حيث تعيش في بيئات متنوعة مثل الأنظمة المائية العذبة والبرية، بالإضافة إلى المواطن القاسية مثل الينابيع الساخنة، والأنظمة المائية عالية الملوحة، والمناطق القطبية، والبيئات ذات الأوكسجين المنخفض. تعتبر البلازما وسيلة فعالة للغاية لتحلل الخلايا. عند تعرض *Nostoc muscorum*، وهو نوع من السيانوبكتيريا، لبلازما الهواء، يتم تدميره تمامًا خلال خمسة عشر دقيقة. بالمقابل، استغرق استخدام بلازما الأوكسجين والنيتروجين ساعة على الأقل لتحقيق نفس النتيجة. ولدت البلازما الجوية أنواعًا كيميائية مثل $N_2^+$، $N^+$، $OH$ بتركيزات أعلى من بلازما الأوكسجين والنيتروجين. تم قياس طيف البلازما أثناء معالجة السيانوبكتيريا، كما تم تقدير درجة حرارة الإلكترون وكثافة البلازما في بلازما الهواء. اختفى امتصاص الأشعة الضوئية لأصباغ السينوبكتيريا والقمم المميزة لطيف الرامان بعد المعالجة بالبلازما، مما يعد دليلاً واضحًا على التحلل التام للسيانوبكتيريا.